\documentclass[prd,aps,tightenlines,a4paper,12pt]{revtex4}

\usepackage{graphicx}
\usepackage{bm}

\usepackage{url}

\newcommand{\OO}[1]{{\mathcal O}(c^{-#1})}

\newcommand{\muas}[0]{\hbox{\rm $\mu$as}}

\newcommand{\ve}[1]{\mbox{\boldmath$#1$}}
\def\source{{\rm 0}}
\def\obs{{\rm 1}}

\begin{document}

\title{Further simplification of the light deflection formula \\
for solar system objects}

\author{Sven \surname{Zschocke}, Sergei A. \surname{Klioner}}

\affiliation{Lohrmann Observatory, Dresden Technical University,
Mommsenstr. 13, 01062 Dresden, Germany}

\begin{abstract}

\begin{center}
{\it GAIA-CA-TN-LO-SZ-005-1}

\medskip

\today

\end{center}

The transformation $\ve{n}$ to $\ve{k}$ in post-post-Newtonian order is simplified. All post-post-Newtonian terms 
of the order ${\cal O} \left(\frac{\displaystyle m^2}{\displaystyle d^2}\right)$ are neglected and we show that the 
total sum of these terms is smaller than 
$\frac{\displaystyle 15}{\displaystyle 4}\,\pi\,\frac{\displaystyle m^2}{\displaystyle d^2}$.
This simpler transformation will improve the efficiency of Gaia data reduction. 

\end{abstract}

\maketitle

\newpage

\section{Introduction}

The approximative analytical solution of the problem of light deflection has been presented in  
\cite{Article_Klioner_Zschocke,Report1,Report2}. One of the main result 
of these investigations is the transformation $\ve{n}$ to $\ve{k}$ for solar system objects in post-post-Newtonian 
approximation. A detailed analysis \cite{Report2,Report3} has shown that most of the terms in this transformation 
can be neglected at the micro-arcsecond level of accuracy, leading a simplified formula $\ve{n}$ 
to $\ve{k}$ for the data reduction. This simplified formula $\ve{n}$ to $\ve{k}$ has been given in Eqs.~(92) and 
(93) in \cite{Article_Klioner_Zschocke} and in Eqs.~(52) and (53) in \cite{Report2}. In this report we will show 
that this transformation can be further simplified. The report is organized as follows. 
In Section \ref{Section_I} we will present the transformation $\ve{n}$ to $\ve{k}$ in 
post-post-Newtonian order. The estimate of post-post-Newtonian terms and the new simplified transformation 
$\ve{n}$ to $\ve{k}$ in given in Section \ref{Section_II}. A new estimation will be given 
in Section \ref{Section_III}. A summary is given in Section \ref{Section_IV}. 
Detailed proofs of the estimates used are given in the appendices. 

\section{Transformation $\ve{n}$ to $\ve{k}$ in post-post-Newtonian order\label{Section_I}}

The transformation $\ve{n}$ to $\ve{k}$ in post-post-Newtonian order has been given in 
Eq.~(87) in \cite{Article_Klioner_Zschocke}, Eq.~(57) in \cite{Report1}, and in Eq.~(45) in \cite{Report2}. 
We will present this transformation in the following equivalent form:

\begin{eqnarray}
{\phantom{\biggr|}}_{\rm N}&\biggr|&\quad
\ve{n} = \ve{k}
\nonumber\\
{\phantom{\biggr|}}_{\rm pN}&\biggr|&
\phantom{\ve{n} = \ve{k}}
 - (1 + \gamma) \, m \,\frac{\ve{k} \times
(\ve{x}_{\source} \times \ve{x}_{\obs})}{x_{\obs}\left(x_{\obs}\,x_{\source} + \ve{x}_{\obs}\cdot\ve{x}_{\source}\right)}
\nonumber\\
{\phantom{\biggr|}}_{\Delta\rm pN}&\biggr|&
\phantom{\ve{n} = \ve{k}}
 + (1 + \gamma)^2 \, m^2 \,\frac{\ve{k} \times
(\ve{x}_{\source} \times \ve{x}_{\obs})}{\left(x_{\obs}\,x_{\source} + \ve{x}_{\obs}\cdot\ve{x}_{\source}\right)^2}
\,{R \over x_{\obs}}
\nonumber\\
{\phantom{\biggr|}}_{\rm scaling}&\biggr|&
\phantom{\ve{n} = \ve{k}}
- \frac{1}{8}\,(1 + \gamma)^2\,\frac{m^2}{x_{\obs}^2}\,\ve{k}\,
\frac{{\left((x_{\obs} - x_{\source})^2 - R^2\right)}^2}{|\ve{x}_{\obs} \times \ve{x}_{\source}|^2}
\nonumber\\
{\phantom{\biggr|}}_{\rm ppN}&\biggr|&
\phantom{\ve{n} = \ve{k}}
 + \,m^2\, \ve{k} \times (\ve{x}_{\source} \times \ve{x}_{\obs})\,
\Biggl[
\,{1\over 2}\,(1 + \gamma)^2\,
\frac{R^2-(x_{\obs}-x_{\source})^2}{x_{\obs}^2\,|\ve{x}_{\obs} \times \ve{x}_{\source}|^2}
\nonumber\\
{\phantom{\biggr|}}_{\rm ppN}&\biggr|&
\phantom{\ve{n} = \ve{k}}
 + \, \frac{1}{4} \, \alpha \, \epsilon \, \frac{1}{R}
\left(\frac{1}{R\,x_{\source}^2} - \frac{1}{R\,x_{\obs}^2}
- 2\, \frac{\ve{k} \cdot \ve{x_{\obs}}}{x_{\obs}^4}\right)
\nonumber\\
{\phantom{\biggr|}}_{\rm ppN}&\biggr|&
\phantom{\ve{n} = \ve{k}}
 - \frac{1}{4}\,\left(\, 8(1 + \gamma - \alpha \gamma) (1 + \gamma) - 4\alpha \beta
+ 3\, \alpha\, \epsilon \, \right)  \, R\,\frac{\ve{k} \cdot \ve{x}_{\obs}}
{x_{\obs}^2\,|\, \ve{x}_{\obs} \times \ve{x}_{\source} \,|^2}
\nonumber\\
{\phantom{\biggr|}}_{\rm ppN}&\biggr|&
\phantom{\ve{n} = \ve{k}}
 + \frac{1}{8}\, \left(8 (1 + \gamma - \alpha \, \gamma) (1 + \gamma)
- 4 \, \alpha \, \beta + 3 \alpha \, \epsilon\right) \,
\frac{x_{\obs}^2 - x_{\source}^2 - R^2}{|{\ve{x}_{\obs}} \times {\ve{x}}_{\source}|^3}
\,\delta(\ve{x}_{\obs} , \ve{x}_{\source})
\Biggr]
\nonumber\\
{\phantom{\biggr|}}_{\rm ppN}&\biggr|&
\phantom{\ve{n} = \ve{k}}
+ (1 + \gamma)^2 \, m^2 \,\frac{\ve{k} \times
(\ve{x}_{\source} \times \ve{x}_{\obs})}{\left(x_{\obs}\,x_{\source} + \ve{x}_{\obs}\cdot\ve{x}_{\source}\right)^2}\,
{x_{\obs} + x_{\source} - R \over x_{\obs}}
\nonumber\\
&& \phantom{\ve{n} = \ve{k}} +\OO6\,.
\label{Eq_5}
\end{eqnarray}
\noindent
Here we have classified the nature of the individual terms by labels N (Newtonian), pN (post-Newtonian), ppN 
(post-post-Newtonian) and $\Delta\rm pN$ (terms that are formally of post-post-Newtonian order, but may numerically 
become significantly larger than other post-post-Newtonian terms, see estimates in (\ref{Eq_50})).

\section{Simplified transformation $\ve{n}$ to $\ve{k}$\label{Section_II}}

The effect of all the ``ppN'' terms in (\ref{Eq_5}) can be estimated as 
(cf. Eq.~(91) in \cite{Article_Klioner_Zschocke} or Eq.~(50) in \cite{Report2})

\begin{eqnarray}
\left|\, \ve{\omega}_{\rm ppN}^{\;\prime} \,\right| &\le& {15\over 4}\,\pi\, {m^2\over d^2}\,.
\label{Eq_30}
\end{eqnarray}

\noindent
The proof of (\ref{Eq_30}) is given in Appendix \ref{Appendix_B}.
These terms can attain $1\,\muas$ only for observations within about $3.3$ angular radii from the Sun and can be
neglected. Accordingly, we obtain a simplified formula for the transformation from $\ve{k}$ to $\ve{n}$ keeping
only the post-Newtonian and ``enhanced'' post-post-Newtonian terms labelled as ``pN'' and ``$\Delta$pN''
in (\ref{Eq_5}):

\begin{eqnarray}
\ve{n} &=& \ve{k}
+\ve{d}\,P\,\left(1+P\,x_{\obs}\right)
+{\cal O}\left({m^2\over d^2}\right)+{\cal O}({m^3})\,,
\label{Eq_35}
\\
P&=&-(1+\gamma)\,{m\over d^2}\,
\left({x_{\source}-x_{\obs}\over R}+{\ve{k}\cdot\ve{x}_{\obs}\over x_{\obs}}\right)\,.
\label{Eq_40}
\end{eqnarray}
\noindent
The simplified transformation $\ve{n}$ to $\ve{k}$ given in Eq.~(\ref{Eq_35}) has now simpler
structure than the former expression given in Eq.~(92) in \cite{Article_Klioner_Zschocke} or in Eq.~(52) in
\cite{Report2}. Therefore, (\ref{Eq_35}) is more efficient for the data reduction. 
Furthermore, the transformation in Eq.~(\ref{Eq_35}) has now similar structure as the
simplified transformation $\ve{n}$ to $\ve{\sigma}$ given in Eq.~(102) in
\cite{Article_Klioner_Zschocke} or in Eq.~(62) in \cite{Report2}.

\section{A new estimation\label{Section_III}}

The enhanced post-post-Newtonian term $\left|\,\ve{\omega}_{\Delta pN}^{\;\prime}\,\right|$ in Eq.~(\ref{Eq_5}) is, 
for $\gamma=1$, given by (cf. Eq.~(89) in \cite{Article_Klioner_Zschocke} or Eq.~(48) in \cite{Report2})

\begin{eqnarray}
\left|\,\ve{\omega}_{\Delta pN}^{\;\prime}\,\right| = 4 \, m^2 \,\frac{\left|\,\ve{k} \times
(\ve{x}_{\source} \times \ve{x}_{\obs})\,\right|}
{\left(x_{\obs}\,x_{\source} + \ve{x}_{\obs}\cdot\ve{x}_{\source}\right)^2}\,{R \over x_{\obs}}\,.
\label{Eq_45}
\end{eqnarray}

\noindent
This term differs from the corresponding term $\left|\,\ve{\omega}_{\Delta pN}\,\right|$ defined in Eq.~(89) in  
\cite{Article_Klioner_Zschocke} or Eq.~(48) in \cite{Report2} only by a factor 
$\frac{\displaystyle R}{\displaystyle x_{\source} + x_{\obs}} \le 1$. Therefore, we conclude that the estimates 
given in Eqs.~(89) and (90) of \cite{Article_Klioner_Zschocke} or in Eqs.~(48) and (49) of \cite{Report2} 
are also valid for $\left|\,\ve{\omega}_{\Delta pN}^{\;\prime}\,\right|$, that means:

\begin{eqnarray}
\left|\,\ve{\omega}_{\Delta pN}^{\;\prime}\,\right| 
&\le& 16\,\frac{m^2}{d^3}\,\frac{R^2\,x_{\obs}\,x_{\source}^2}{\left(x_{\obs} + x_{\source}\right)^4}
\le 16\,\frac{m^2}{d^3}\,\frac{R\,x_{\obs}\,x_{\source}^2}{\left(x_{\obs} + x_{\source}\right)^3} 
\le 16\,\frac{m^2}{d^3}\,\frac{x_{\obs}\,x_{\source}^2}{\left(x_{\obs} + x_{\source}\right)^2}
\le 16\,\frac{m^2}{d^2}\,\frac{x_{\obs}}{d}\,,
\label{Eq_50}
\end{eqnarray}

\noindent
where the first expression given in (\ref{Eq_50}) represents a new estimation. Another estimation can be given, 
namely (cf. Eq.~(90) in \cite{Article_Klioner_Zschocke} or Eq.~(49) in \cite{Report2}) 

\begin{eqnarray}
|\,\ve{\omega}_{\Delta\rm pN}^{\;\prime}\,| &\le& {64\over 27}\,\frac{m^2}{d^2}\,{R\over d}\,,
\label{Eq_55}
\end{eqnarray}

\noindent
which cannot be related to the estimations in (\ref{Eq_50}) and reflect different properties of
$\left|\,\ve{\omega}_{\Delta\rm pN}^{\;\prime}\,\right|$ as function of multiple variables.

\section{Summary\label{Section_IV}}

In Eq.~(57) in \cite{Report1} the complete transformation $\ve{n}$ to $\ve{k}$ in post-post-Newtonian 
order has been given. In \cite{Report2} we have shown that most of the terms can be neglected 
because they are of the order ${\cal O} \left(\frac{\displaystyle m^2}{\displaystyle d^2}\right)$ 
and can attain $1\,\muas$ only for observations within about $3.3$ angular radii from the Sun. 
These investigations have yielded a simplified transformation, given in Eqs.~(92) and (93) in 
\cite{Article_Klioner_Zschocke} or in Eqs.~(52) and (53) in \cite{Report2}, and applicable for an efficient 
data reduction. In this report we have shown that Eq.~(92) in \cite{Article_Klioner_Zschocke} or Eq.~(52) in 
\cite{Report2} can further be simplified. The main result of this report is Eq.~(\ref{Eq_35}), where we give a 
new simplified transformation $\ve{n}$ to $\ve{k}$ which will improve the efficiency of Gaia data reduction. 
We have shown that the total sum of the neglected ppN-terms is smaller than 
$\frac{\displaystyle 15}{\displaystyle 4}\,\pi\,\frac{\displaystyle m^2}{\displaystyle d^2}$. Furthermore, 
estimations of the enhanced post-post-Newtonian term has been given in Eqs.~(\ref{Eq_50}) and (\ref{Eq_55}).

\newpage

\appendix

\section{Proof of inequality (\ref{Eq_30})\label{Appendix_B}}

The sum of all ppN-terms in Eq.~(\ref{Eq_5}) can be written as follows (here $\alpha=\beta=\gamma=\epsilon=1$):

\begin{eqnarray}
\left|\,\ve{\omega}_{\rm ppN}^{\;\prime}\,\right| &=& \frac{1}{4}\,\frac{m^2}{d^2}\,f_{10}^{\;\prime}\,,
\label{appendix_B_5}
\end{eqnarray}

\noindent
where the function is defined by (cf. with $f_{10}$ defined in Eq.~(84) in \cite{Report3}) 

\begin{eqnarray}
f_{10}^{\;\prime} &=& \Bigg|\,{z\,(16z-z\,\cos\Phi-15)\,\sin\Phi\over 1+z^2-2z\,\cos\Phi}
+ \,{z(1-3z^2+2z^3\cos\Phi)\,\sin^3\Phi\over \left(1+z^2-2z\,\cos\Phi\right)^2}
\nonumber\\
\nonumber\\
&& + \,{15z\,(\cos\Phi-z)\,\Phi\over 1+z^2-2z\,\cos\Phi}
+ 16\, \frac{z\,\left(1 - \cos \Phi\right)^2\,\left(1 + z - \sqrt{1 + z^2 - 2\,z\,\cos \Phi}\right)}
{\left(1+z^2 - 2\,z\,\cos \Phi\right)\,\sin \Phi}\,\Bigg|\,.
\label{appendix_B_10}
\end{eqnarray}

\noindent
Here we have used the notation $\Phi = \delta \left(\ve{x}_{\source}, \ve{x}_{\obs}\right)$ 
and $z = \frac{\displaystyle x_{\source}}{\displaystyle x_{\obs}}$. By means of the inequalities 
(note that (\ref{appendix_B_20}) improves the inequality given in Eq.~(C1) in \cite{Report3})

\begin{eqnarray}
f_2 &=& 16\, \frac{z\,\left(1 - \cos \Phi\right)^2\,\left(1 + z - \sqrt{1 + z^2 - 2\,z\,\cos \Phi}\right)}
{\left(1+z^2 - 2\,z\,\cos \Phi\right)\,\sin \Phi} \,\le\, 8\,\sin \Phi\,,
\label{appendix_B_15}
\\
\nonumber\\
f_3 &=& {\left|\,z\,\left(1-3z^2+2z^3\cos\Phi\right)\,\right|\,\sin^3\Phi \over \left(1+z^2-2z\,\cos\Phi\right)^2} 
\,\le\, 3\,\sin \Phi\,,
\label{appendix_B_20}
\end{eqnarray}

\noindent
(proof of (\ref{appendix_B_15}) and (\ref{appendix_B_20}) are shown in Appendices \ref{Appendix_F} and 
\ref{Appendix_E}, respectively) we obtain 

\begin{eqnarray}
f_{10}^{\;\prime} &\le& 
\left|\,{z\,(16z-z\,\cos\Phi-15)\,\sin\Phi\over 1+z^2-2z\,\cos\Phi}
+ {15z\,(\cos\Phi-z)\,\Phi\over 1+z^2-2z\,\cos\Phi}\,\right| + 11\,\sin \Phi\,.
\label{appendix_B_25}
\end{eqnarray}

\noindent
In \cite{Report3} we have shown $z\,(16z-z\,\cos\Phi-15)\,\sin\Phi + 15z\,(\cos\Phi-z)\,\Phi \le 0$.
Accordingly, due to $\sin \Phi \ge 0$, we obtain 

\begin{eqnarray}
f_{10}^{\;\prime} &\le& 
\left|\,{z\,(16z-z\,\cos\Phi-15)\,\sin\Phi\over 1+z^2-2z\,\cos\Phi}
+ {15z\,(\cos\Phi-z)\,\Phi\over 1+z^2-2z\,\cos\Phi} - 15\,\sin \Phi\,\right|\,,
\label{appendix_B_35}
\end{eqnarray}

\noindent
where, for convenience, we have replaced the term $11\,\sin \Phi$ by the larger term $15\,\sin \Phi$. 
Furthermore, in \cite{Report3} we have shown that 

\begin{eqnarray}
\left|\,{z\,(16z-z\,\cos\Phi-15)\,\sin\Phi\over 1+z^2-2z\,\cos\Phi}
+ {15z\,(\cos\Phi-z)\,\Phi\over 1+z^2-2z\,\cos\Phi} - 15\,\sin \Phi\,\right| &\le& 15\,\pi\,.
\label{appendix_B_40}
\end{eqnarray}

\noindent
Thus, we obtain 

\begin{eqnarray}
f_{10}^{\;\prime} &\le& 15\,\pi\,.
\label{appendix_B_45}
\end{eqnarray}

\noindent
The inequality (\ref{appendix_B_45}) in combination with (\ref{appendix_B_5}) shows the validity of 
inequality (\ref{Eq_30}).

\newpage

\section{Proof of inequalities (\ref{appendix_B_15}) \label{Appendix_F}}

In order to show (\ref{appendix_B_15}), we rewrite this inequality as follows:

\begin{eqnarray}
\frac{z\,\left(1 - \cos \Phi \right)}{1 + z^2 - 2\,z\,\cos \Phi}\,
\frac{1 + z - \sqrt{1+z^2-2\,z\,\cos \Phi}}{1 + \cos \Phi} &\le& \frac{1}{2}\,.
\label{appendix_E_20}
\end{eqnarray}

\noindent
The inequality (\ref{appendix_E_20}) can be splitted into two factors satisfying the following inequalities:

\begin{eqnarray}
\frac{z\,\left(1 - \cos \Phi \right)}{1+z^2-2\,z\,\cos \Phi} &\le& \frac{1}{2}\,,
\label{appendix_E_25}
\\
\nonumber\\
\frac{1 + z - \sqrt{1+z^2-2\,z\,\cos \Phi}}{1+\cos \Phi} &\le& 1\,.
\label{appendix_E_30}
\end{eqnarray}

\noindent
The inequality (\ref{appendix_E_25}) is obviously valid, because by multiplying (\ref{appendix_E_25}) with the 
denominator we obtain $- \left( 1 - z\right)^2 \le 0$. 
The inequality (\ref{appendix_E_30}) is also straightforward, because it 
can be rewritten as $z - \cos \Phi \le \sqrt{1+z^2 - 2\,z\,\cos \Phi}$, which is obviously valid due to 
$z - \cos \Phi \le |\,z - \cos \Phi \,|$. Thus we have shown the validity of inequality (\ref{appendix_E_20}) 
and (\ref{appendix_B_15}), respectively.

\newpage

\section{Proof of inequality (\ref{appendix_B_20}) \label{Appendix_E}}

Using the notation $w = \cos \Phi$, the inequality (\ref{appendix_B_20}) can be written as follows:

\begin{eqnarray}
f_3 &=& \frac{z\,\left|\,1 - 3\,z^2 + 2\,z^3 \,w\,\right|\,\left(1 - w^2 \right)}{\left(1 + z^2 - 2\,w\,z\right)^2} \le 3\,.
\label{appendix_5}
\end{eqnarray}

\noindent
Using the inequality (proof see below)

\begin{eqnarray}
\left|\,1 - 3\,z^2 + 2\,z^3 \,w\,\right| &\le& 1 - 3\,w\,z^2 + 2\,z^3\,,
\label{appendix_10}
\end{eqnarray}

\noindent
we obtain 

\begin{eqnarray}
f_3 &\le& \frac{z\,\left(1 - 3\,w\,z^2 + 2\,z^3\right)\,\left(1 - w^2 \right)}{\left(1 + z^2 - 2\,w\,z\right)^2} 
= h_1 + h_2 \le 3\,.
\label{appendix_15}
\end{eqnarray}

\noindent
In (\ref{appendix_15}) the relation 
$1 - 3\,w\,z^2 + 2\,z^3 = \left( 1 + z^2 - 2\,w\,z \right) +\left( - 3\,w\,z^2 + 2\,z^3 -z^2 + 2\,w\,z \right)$ 
has been used. The functions are defined by 

\begin{eqnarray}
h_1 &=& \frac{z\,\left(1 - w^2\right)}{1 + z^2 - 2\,w\,z} \le \frac{2\,z\,\left(1-w\right)}{1+z^2-2\,w\,z}\le 1\,,
\label{appendix_25}
\\
\nonumber\\
h_2 &=& \frac{z^2\,\left|\,-3\,w\,z + 2\,z^2 - z + 2\,w\,\right|\,\left(1- w^2\right)}{\left(1+z^2-2\,w\,z\right)^2}\le 2\,.
\label{appendix_30}
\end{eqnarray}

\noindent
The inequality (\ref{appendix_25}) has been shown in \cite{Report3}. 
In order to show (\ref{appendix_30}), we factorize the function $h_2$ as follows:

\begin{eqnarray}
h_2 &=& h_2^A\;h_2^B\,,
\label{appendix_35}
\\
\nonumber\\
h_2^A &=& \frac{z^2\,\left(1 - w^2\right)}{1 + z^2 - 2\,w\,z} \le 1\,,
\label{appendix_40}
\\
\nonumber\\
h_2^B &=& \frac{\left|\,- 3\,w\,z + 2\,z^2 - z + 2\,w\,\right|}{1 + z^2 - 2\,w\,z} \le 2\,.
\label{appendix_45}
\end{eqnarray}

\noindent
Thus, by means of the inequalities (\ref{appendix_10}) and (\ref{appendix_25}) - (\ref{appendix_45}) 
we have shown the validity of inequality (\ref{appendix_5}) and (\ref{appendix_B_20}), respectively. 
We still have to proof of inequalities (\ref{appendix_10}), (\ref{appendix_40}) and (\ref{appendix_45}). 

Let us consider (\ref{appendix_10}). First, we remark that $1 - 3\,w\,z^2 + 2\,z^3 \ge 0$ because of 
$1 - 3\,z^2 + 2\,z^3 \ge 0$. Then, squaring both sides of (\ref{appendix_10}) and subtracting from each other 
leads to 

\begin{eqnarray}
h_3 &=& 2\,z^3 + 2\,w\,z^3 - 3\,z^2 - 3\,w\,z^2 + 2 \ge 0\,.
\label{appendix_50}
\end{eqnarray}

\noindent
The boundaries of $h_3$ are 

\begin{eqnarray}
\lim_{w \rightarrow - 1} h_3 &=& 2 \ge 0\,,
\label{appendix_55}
\\
\nonumber\\
\lim_{w \rightarrow + 1} h_3 &=& 2\,\left(2\,z+1\right)\,\left(z - 1 \right)^2 \ge 0\,,
\label{appendix_60}
\\
\nonumber\\
\lim_{z \rightarrow 0} h_3 &=& 2 \ge 0\,,
\label{appendix_65}
\\
\nonumber\\
\lim_{z \rightarrow \infty} h_3 &=& 2\,\left(1 + w \right)\,\lim_{z \rightarrow \infty} z^3 \ge 0\,.
\label{appendix_70}
\end{eqnarray}

\noindent
The extremal conditions $h_{3\,,\,w} = 0$ and $h_{3\,,\,z} = 0$ lead to 

\begin{eqnarray}
z^2\,\left(2\,z - 3\right) &=& 0\,,
\label{appendix_75}
\\
\nonumber\\
z\,\left(1 + w\right)\,\left(z - 1 \right) &=& 0\,.
\label{appendix_80}
\end{eqnarray}

\noindent
The common solutions of (\ref{appendix_75}) and (\ref{appendix_80}) are given by 

\begin{eqnarray}
P_1 \left(w=-1\,,\,z=0\right)\,,
\label{appendix_85}
\\
\nonumber\\
P_2 \left(w=-1\,,\,z=\frac{3}{2}\right)\,.
\label{appendix_90}
\end{eqnarray}

\noindent
The numerical values of $h_3$ at these turning points are 

\begin{eqnarray}
h_3 \left(P_1\right) &=& 2 \ge 0\,,
\label{appendix_95}
\\
\nonumber\\
h_3 \left(P_2\right) &=& 2 \ge 0\,.
\label{appendix_100}
\end{eqnarray}

\noindent
Thus, we have shown (\ref{appendix_50}) and, therefore, the validity of inequality (\ref{appendix_10}).

Now we consider the inequality (\ref{appendix_40}). Multiplying both sides of this relation with the denominator 
leads to the inequality 

\begin{eqnarray}
h_4 &=& - z^2\,w^2 - 1 + 2\,w\,z \le 0\,.
\label{appendix_105}
\end{eqnarray}

\noindent
The boundaries of $h_4$ are 

\begin{eqnarray}
\lim_{w \rightarrow - 1} h_4 &=& - \left(1 + z \right)^2 \le 0\,,
\label{appendix_110}
\\
\nonumber\\
\lim_{w \rightarrow + 1} h_4 &=& - \left(1 - z \right)^2 \le 0\,,
\label{appendix_115}
\\
\nonumber\\
\lim_{z \rightarrow 0} h_4 &=& - 1 \le 0\,,
\label{appendix_120}
\\
\nonumber\\
\lim_{z \rightarrow \infty} h_4 &=& - w^2\,\lim_{z \rightarrow \infty} z^2 \le 0\,.
\label{appendix_125}
\end{eqnarray}

\noindent
The extremal conditions $h_{4\,,\,w} = 0$ and $h_{4\,,\,z} = 0$ lead to

\begin{eqnarray}
z\,\left( 1 - w\,z\right) &=& 0\,,
\label{appendix_130}
\\
\nonumber\\
w\,\left( 1 - w\,z\right) &=& 0 \,.
\label{appendix_135}
\end{eqnarray}

\noindent
The common solution of (\ref{appendix_130}) and (\ref{appendix_135}) is given by

\begin{eqnarray}
P_3 \left(w = 0 \,,\,z=0\right)\,,
\label{appendix_140}
\end{eqnarray}

\noindent
and the numerical value of $h_4$ at this turning point is

\begin{eqnarray}
h_4 \left(P_3\right) &=&  - 1 \le 0 \,.
\label{appendix_145}
\end{eqnarray}

\noindent
Thus, we have shown (\ref{appendix_105}) and, therefore, the inequality (\ref{appendix_40}).

Now we consider the inequality (\ref{appendix_45}). Squaring both sides of (\ref{appendix_45}) 
and subtracting from each other leads to the inequality 

\begin{eqnarray}
h_5 &=& 4\,z^2 - z - 7\,w\,z + 2 + 2\,w \ge 0\,.
\label{appendix_150}
\end{eqnarray}

\noindent
The boundaries of $h_5$ are

\begin{eqnarray}
\lim_{w \rightarrow - 1} h_5 &=& 2\,z \,\left(3 + 2\,z\right) \ge 0\,,
\label{appendix_155}
\\
\nonumber\\
\lim_{w \rightarrow + 1} h_5 &=& 4\,\left(z - 1\right)^2 \ge 0\,,
\label{appendix_160}
\\
\nonumber\\
\lim_{z \rightarrow 0} h_5 &=& 2 \left( 1+ w \right) \ge 0\,,
\label{appendix_165}
\\
\nonumber\\
\lim_{z \rightarrow \infty} h_5 &=& 4\,\lim_{z \rightarrow \infty} z^2 \ge 0\,.
\label{appendix_170}
\end{eqnarray}

\noindent
The extremal conditions $h_{5\;,w} = 0$ and $h_{5\;z} = 0$ lead to

\begin{eqnarray}
-7\,z + 2 &=& 0\,,
\label{appendix_175}
\\
\nonumber\\
8\,z - 1 - 7\,w &=& 0\,.
\label{appendix_180}
\end{eqnarray}

\noindent
The common solution of (\ref{appendix_175}) and (\ref{appendix_180}) is given by

\begin{eqnarray}
P_4 \left(w = \frac{9}{49} \,,\,z=\frac{2}{7}\right)\,,
\label{appendix_185}
\end{eqnarray}

\noindent
and the numerical value of $h_5$ at this turning point is

\begin{eqnarray}
h_5 \left(P_4\right) &=&  \frac{100}{49} \ge 0 \,.
\label{appendix_190}
\end{eqnarray}

\noindent
Thus, we have shown (\ref{appendix_150}) and, therefore, the inequality (\ref{appendix_45}).

\end{document}